\documentclass[shortnote]{jpsj2} 
\title{$^{75}$As NMR Observation of Anisotropic Spin Fluctuations in the Ba(Fe$_{0.92}$Co$_{0.08}$)$_2$As$_2$ Superconductor (T$_c$ = 22 K)}

\author{Fanlong \textsc{Ning}$^{1}$, Kanagasingham \textsc{Ahilan}$^{1}$, Takashi \textsc{Imai}$^{1,2}$, Athena S. \textsc{Sefat}$^{3}$, Rongying \textsc{Jin}$^{3}$, Michael A. \textsc{Mcguire}$^{3}$, Brian C. \textsc{Sales}$^{3}$, David \textsc{Mandrus}$^{3}$}

\inst{$^{1}$Department of Physics and Astronomy, McMaster
University, Hamilton, Ontario L8S4M1, Canada \\
$^{2}$Canadian Institute for Advanced Research,
Toronto, Ontario M5G1Z8, Canada \\
$^{3}$Materials Science and Technology Division, Oak Ridge National
Laboratory, TN 37831, USA}

\kword{Ba(Fe$_{1-x}$Co$_{x}$)$_2$As$_2$, NMR, anisotropic spin
fluctuations}

\begin{document}
\maketitle

The recent discovery of iron-pnictide superconductors with
transition temperatures as high as T$_c$$\sim$ 55 K has attracted a
huge amount of attention \cite{Kamihara, RenZA,Rotter, Athena1}. A
prototypical parent compound of the iron-pnictide superconductors,
BaFe$_2$As$_2$, is an itinerant antiferromagnet, and displays a
first order Spin Density Wave (SDW) transition at T$_{SDW}$ $\sim$
135 K accompanied by a simultaneous tetragonal-orthorhombic
structural phase transition\cite{Rotter2,HuangQ}. In a recent paper
\cite{Takigawa}, Kitagawa et al. reported a comprehensive $^{75}$As
NMR investigation of the parent compound BaFe$_2$As$_2$, and
proposed a stripe antiferromagnetic spin structure below T$_{SDW}$.
They also showed the growth of anisotropic spin fluctuations at
$^{75}$As sites between $\textit{ab}$- and $\textit{c}$-axis
orientations in the paramagnetic state above T$_{SDW}$. In this
short note, we will demonstrate that analogous anisotropy of
paramagnetic spin fluctuations grows with decreasing temperature in
the optimally electron-doped superconductor
Ba(Fe$_{0.92}$Co$_{0.08}$)$_2$As$_2$ (T$_c$ = 22 K) when short-range
antiferromagnetic correlations develop toward T$_c$.

\begin{figure}[!htpb]\vspace{-0.5cm}
\centering\includegraphics[width = 10cm, angle =0]{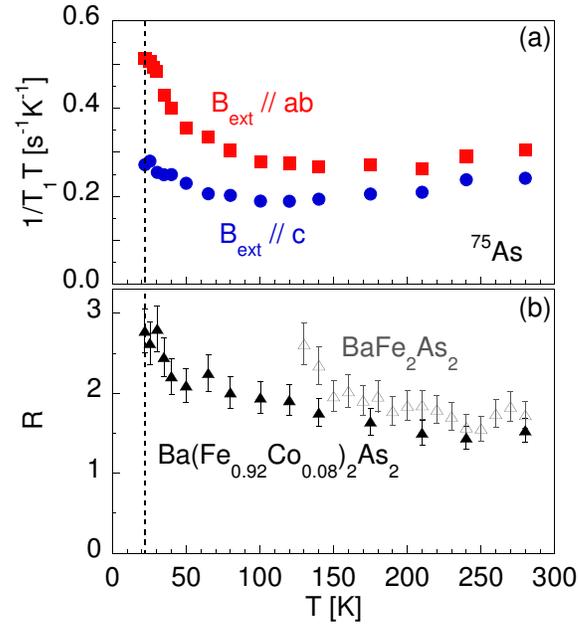}\vspace{-3.5cm}
 \caption{ (Color Online). (a) The temperature dependence of $^{75}(\frac{1}{T_1T})$ with
external field B$_{ext}$ = 7.7 Tesla applied within the
$\textit{ab}$-plane ($\blacksquare$) and along the $\textit{c}$-axis
($\bullet$) in Ba(Fe$_{0.92}$Co$_{0.08}$)$_2$As$_2$. (b) The
temperature dependence of the anisotropy of spin fluctuations R for
the superconductor Ba(Fe$_{0.92}$Co$_{0.08}$)$_2$As$_2$
($\diamond$), and BaFe$_2$As$_2$ ($\triangle$) (estimated from [7],
T$_{SDW}$ = 135 K). The dashed line marks T$_c$.}
\label{fig:Ning_fig1}
\end{figure}

In Fig. \ref{fig:Ning_fig1}(a), we show the $^{75}$As nuclear
spin-lattice relaxation rate divided by temperature,
$^{75}(\frac{1}{T_1T})_{ab}$, for
Ba(Fe$_{0.92}$Co$_{0.08}$)$_2$As$_2$ with an external magnetic field
B$_{ext}$ = 7.7 Tesla applied within the $\textit{ab}$-plane. For
comparison, we also reproduce $^{75}(\frac{1}{T_1T})_{c}$ with
B$_{ext}$ applied along the $\textit{c}$-axis from our earlier
report \cite{Ning1}. Compared with the case of BaFe$_2$As$_2$ in
[7], the overall magnitude of $^{75}(\frac{1}{T_1T})$ in the
paramagnetic state is suppressed by Co doping for both orientations.
We refer readers to our previous studies \cite{Ning1,Ning2} for a
systematic investigation of the Co doping effects on the
static and dynamic susceptibilities. Our results in Fig. \ref{fig:Ning_fig1}(a) show
that the enhancement of $(\frac{1}{T_1T})$ toward T$_c$ is stronger
for $^{75}(\frac{1}{T_1T})_{ab}$ than for $^{75}(\frac{1}{T_1T})_c$.
This behavior in Ba(Fe$_{0.92}$Co$_{0.08}$)$_2$As$_2$ is
qualitatively similar to that of the undoped parent compound
BaFe$_2$As$_2$ above T$_{SDW}$ \cite{Takigawa}.

In general, $(\frac{1}{T_1T})$ probes spin fluctuations within the
plane perpendicular to the quantization axis set by B$_{ext}$,

\begin{subequations}
\begin{eqnarray}
 ({\frac{1}{T_{1}T})_c} &\propto& \sum_{{\bf
q}}{|A_a(\bf{q})|^{2}\frac{\chi_a"({\bf q}, \textit{f
})}{\textit{f}}} + \sum_{{\bf
q}}{|A_{b}(\bf{q})|^{2}\frac{\chi_b"({\bf q}, \textit{f
})}{\textit{f}}}\\
&\propto& 2\sum_{{\bf q}}{|A_{a}(\bf{q})|^{2}\frac{\chi_a"({\bf q},
\textit{f
})}{\textit{f}}},\\
({\frac{1}{T_{1}T})_{ab}} &\propto& \sum_{{\bf
q}}{|A_{a}(\bf{q})|^{2}\frac{\chi_a"({\bf q}, \textit{f
})}{\textit{f}}} + \sum_{{\bf
q}}{|A_{c}(\bf{q})|^{2}\frac{\chi_c"({\bf q}, \textit{f
})}{\textit{f}}},
\end{eqnarray}
\end{subequations}
where $|A_{\alpha}(\bf{q})|^{2}$ ($\alpha$ = a, b, c) and
$\chi_{\alpha}"({\bf q}, \textit{f })$ are the wave-vector
$\bf{q}$-dependent hyperfine form factor and the dynamical spin
susceptibility at NMR frequency $\textit{f}$ , respectively. We
assume $|A_{\alpha}(\bf{q})|^{2}$ and $\chi_\alpha"({\bf q},
\textit{f })$ are isotropic within the $\textit{ab}$-plane in the
tetragonal phase. Thus stronger enhancement of
$^{75}(\frac{1}{T_1T})_{ab}$ toward T$_c$ indicates that spin
fluctuations are enhanced at $^{75}$As sites more strongly toward T$_c$ along the
$\textit{c}$-axis than within the $\textit{ab}$-plane. We define the
anisotropy, R, from Eqs. (1b) and (1c) as,

\begin{equation}
R = \frac{\sum_{{\bf q}}|A_{c}(\bf{q})|^{2}\chi_c"{(\bf q,\textit{f
})}}{\sum_{{\bf q}}|A_{a}(\bf{q})|^{2}\chi_a"{(\bf q,\textit{f })}}
= \frac{(\frac{1}{T_{1}T})_{ab} -
\frac{1}{2}(\frac{1}{T_{1}T})_c}{\frac{1}{2}(\frac{1}{T_{1}T})_c}.
\end{equation}We plot R in Fig. \ref{fig:Ning_fig1}(b). R continuously increases
with decreasing temperature toward T$_c$. Since the magnitudes of
the hyperfine form factors are comparable \cite{Takigawa} and
temperature independent, the observed temperature dependence of R
reflects that of the spin fluctuations. For comparison, we also
estimate R for the parent compound from the data reported in [7] as
shown in Fig. \ref{fig:Ning_fig1}(b). The magnitudes of R for the
two compounds are comparable toward the ordering temperatures. The
anisotropy R below T$_c$ can, in principle, provide valuable
information about the pairing state \cite{Takigawa1}, but it is
beyond the scope of this short note.

\begin{figure}[!htpb]\vspace{-2.5cm}
\centering\includegraphics[width = 10cm, angle =0]{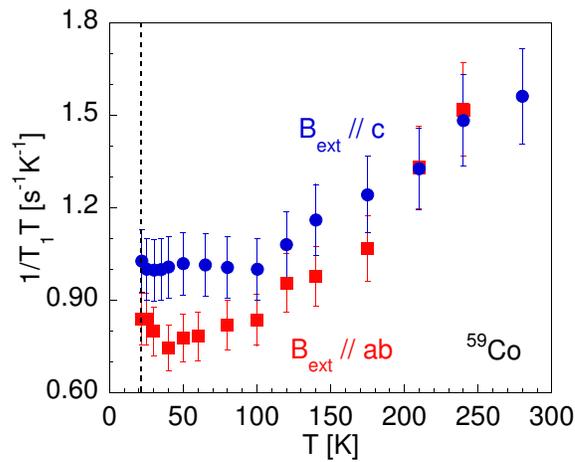}\vspace{-3cm}
 \caption{ (Color Online). $^{59}(\frac{1}{T_1T})$ with
external field B$_{ext}$ applied within the $\textit{ab}$-plane
($\blacksquare$) and along the $\textit{c}$-axis ($\bullet$) for the
superconductor Ba(Fe$_{0.92}$Co$_{0.08}$)$_2$As$_2$. The dashed line
marks T$_c$.} \label{fig:Ning_fig2}
\end{figure}

In Fig. \ref{fig:Ning_fig2}, we present $^{59}(\frac{1}{T_1T})_{ab}$
for Ba(Fe$_{0.92}$Co$_{0.08}$)$_2$As$_2$ with B$_{ext}$ applied
within the $\textit{ab}$-plane. We also present
$^{59}(\frac{1}{T_1T})_c$ for comparison \cite{Ning1}.
$^{59}(\frac{1}{T_1T})_{c}$ levels off below $\sim$ 100 K down to
T$_c$, while $^{59}(\frac{1}{T_1T})_{ab}$ is enhanced slightly
toward T$_c$. This finding provides additional proof that the spin
fluctuations are anisotropic. The presence of the orbital
contributions to $\frac{1}{T_1T}$ \cite{Yafet} of $^{59}$Co makes it difficult to
estimate R. The difference between the temperature
dependences of $^{75}(\frac{1}{T_1T})$ and $^{59}(\frac{1}{T_1T})$
suggest that different areas in the Brillouin Zone have a
different temperature dependence for spin fluctuations, and that the
hyperfine form factors filter out the different regions for
$^{75}$As and $^{59}$Co. Alternatively, because of the itinerant
nature of electrons, the local spin density at $^{75}$As and
$^{59}$Co sites may display different behaviors.

In conclusion, we have demonstrated that the anisotropy R of the
paramagnetic spin fluctuations grows toward T$_c$ at $^{75}$As sites
in the superconductor Ba(Fe$_{0.92}$Co$_{0.08}$)$_2$As$_2$, with
stronger spin fluctuations along the $\textit{c}$-axis. Our finding
is in remarkable contrast with the case of high T$_c$ cuprates,
where R is independent of temperature above T$_c$ \cite{Takigawa1}.

The work at McMaster was supported by NSERC, CIFAR and CFI. Research
at ORNL sponsored by Division of Materials Sciences and Engineering,
Office of Basic Energy Sciences, U.S. Department of Energy.

\end{document}